\documentclass[11pt,a4paper,twoside,reqno,nosumlimits]{amsart}

\usepackage{graphicx}
\usepackage{epstopdf}
\usepackage{amsmath}
\usepackage{amsfonts}
\usepackage{bbm}
\usepackage{braket, url}

\usepackage[normalem]{ulem}

\usepackage{color}
\usepackage{subfigure}
\usepackage[margin=2.0cm]{geometry}
\makeatletter

\newcommand{\Rmnum}[1]{\expandafter\@slowromancap\romannumeral #1@}

\makeatother

\begin{document}
\title{Simplicity bias, algorithmic probability, and the random logistic map}

\author{Boumediene Hamzi$^{1,2}$, Kamaludin Dingle$^{3}$}

 \address{$^1$Department of Computing and Mathematical Sciences, Caltech, CA, USA.}
\address{$^2$The Alan Turing Institute, London, UK.}
\email{boumediene.hamzi@gmail.com}

\address{$^3$Centre for Applied Mathematics and Bioinformatics,
Department of Mathematics and Natural Sciences, 
Gulf University for Science and Technology, Kuwait}
\email{dingle.k@gust.edu.kw}

\date{\today}

\begin{abstract}

\emph{Simplicity bias} is an intriguing phenomenon prevalent in various input-output maps, characterized by a preference for simpler, more regular, or symmetric outputs. Notably, these maps typically feature high-probability outputs with simple patterns, whereas complex patterns are exponentially less probable. This bias has been extensively examined and attributed to principles derived from algorithmic information theory and algorithmic probability. In a significant advancement, it has been demonstrated that the renowned logistic map $x_{k+1}=\mu x_k(1-x_k)$, a staple in dynamical systems theory, and other one-dimensional maps exhibit simplicity bias when conceptualized as input-output systems. Building upon this  work, our research delves into the manifestations of simplicity bias within the random logistic map, specifically focusing on scenarios involving additive noise. This investigation is driven by the overarching goal of formulating a comprehensive theory for the prediction and analysis of time series.

Our primary contributions are multifaceted. We discover that simplicity bias is observable in the random logistic map for specific ranges of $\mu$ and noise magnitudes. Additionally, we find that this bias persists even with the introduction of small measurement noise, though it diminishes as noise levels increase. Our studies also revisit the phenomenon of noise-induced chaos, particularly when $\mu=3.83$, revealing its characteristics through complexity-probability plots. Intriguingly, we employ the logistic map to illustrate a paradoxical aspect of data analysis: more data adhering to a consistent trend can occasionally lead to \emph{reduced} confidence in extrapolation predictions, challenging conventional wisdom.

We propose that adopting a probability-complexity perspective in analyzing dynamical systems could significantly enrich statistical learning theories related to series prediction and analysis. This approach not only facilitates a deeper understanding of simplicity bias and its implications but also paves the way for novel methodologies in forecasting complex systems behavior, especially in scenarios dominated by uncertainty and stochasticity.

\noindent
\emph{Keywords:} Random dynamical systems; algorithmic probability; simplicity bias; time series; machine learning
\end{abstract}

\maketitle

\section{Introduction}

Dynamical systems appear in many fields including financial, engineering, meteorological, medical, and other areas. Because of this, developing methods to predict or understand the behavior and trajectories of dynamical systems is an important area of applied mathematics research. Historically, such predictions have been based on modeling the relevant system (e.g., with ODEs). More recently, however, machine learning advances have given tools for the prediction and analysis of dynamical systems. The machine learning approach instead relies on directly extracting patterns from data rather than detailed modeling of the underlying mechanics \cite{5706920,,yk4, survey_kf_ann, Sindy,jaideep1,nielsen2019practical,   
 kaptanoglu2021physicsconstrained,kutz2022parsimony,   bhcm1,lyap_bh,BHPhysicaD,hamzi2019kernel, bh2020b,klus2020data,ALEXANDER2020132520,bhks,bh12,bh17,hb17,mmd_kernels_bh}. 

It follows that an interesting question is the extent to which machine learning methods apply to dynamical systems analysis and prediction. Because information theory and machine learning are very closely connected \cite{MacKay2003}, this motivates exploring the interface between information theory and dynamical systems research. With this in mind, here we expand on the interface between \emph{algorithmic information theory} \cite{solomonoff1960preliminary,kolmogorov1965three,chaitin1975theory} (AIT), and specifically \emph{algorithmic probability} \cite{solomonoff1960preliminary,solomonoff1964formal,levin1974laws,hutter2007algorithmic}, and dynamical systems. Our current study builds on the work of Ref.\ \cite{dingle2024exploring}, in which various 1D maps from chaos theory were numerically examined from these perspectives.

Random dynamical systems have been investigated for some decades \cite{arnold1995random}. Initially, the field of dynamical systems began with works on deterministic systems, but following this randomly perturbed dynamics became a focus area, and indeed random dynamical systems are a natural way to model physical and other systems (e.g., \cite{dingle2012knudsen}).  Random dynamical systems also provide a rich area of mathematics for study, and they can yield behaviors that are qualitatively different from deterministic dynamical systems, e.g. via noise-induced chaos \cite{mayer1981influence}.

Numerous studies of \emph{simplicity bias} have been made in input-output maps, in which a general inverse relationship between the complexity of outputs and their respective probabilities has been observed \cite{dingle2018input,dingle2020generic}. This work is based on information theory and computation results and was derived from arguments inspired by AIT, and can be viewed as an attempt to apply algorithmic probability in real-world settings \cite{dingle2018input}. The simplicity bias upper bound \cite{dingle2018input} gives a way to bound the probability $P(x)$ of observing output pattern $x$ based essentially only on estimating the Kolmogorov complexity of the pattern in $x$. Simplicity bias analysis has been applied in many contexts, including machine learning and deep neural networks \cite{valle2018deep,mingard2019neural,bhattamishra2022simplicity,yang2019fine,hernandez2021algorithmic,dingle2023multiclass,mingard2023deep}. Simplicity bias and algorithmic probability are closely related to Occam's razor \cite{li2008introduction}, a fundamental basis of scientific reasoning and model selection, that simpler theories/models should be preferred over more complex theories/models provided they explain the data equally well.

The goal of this paper is to bridge the gap between machine learning and algorithmic information theory using dynamical systems, particularly focusing on the concept of simplicity bias within random logistic maps. We aim to extend the understanding of simplicity bias to encompass the inherently stochastic nature of real-world systems as modeled by random dynamical systems. Our approach is twofold: first, to elucidate the manifestation of simplicity bias in the random logistic map under various conditions and noise levels, thereby providing insights into the behavior of these systems; second, to leverage these insights to improve the theoretical foundations and practical applications of machine learning in analyzing and predicting dynamical systems.

We posit that the intersection of dynamical systems, algorithmic information theory, and machine learning represents a fertile ground for advancing our understanding of complex systems. By studying the random logistic map — a prototypical example in dynamical systems theory — we aim to uncover new principles and methods that can inform both the development of machine learning algorithms and the theoretical underpinnings of algorithmic information theory. In doing so, we hope to provide a more robust framework for series prediction and analysis, particularly in contexts where data exhibit complex, chaotic, or noisy behavior.

Ultimately, this paper seeks to not only advance the theoretical discourse in these domains but also to provide a robust set of tools and insights that can be employed in practical applications. Whether it be in financial modeling, climate prediction, or understanding biological systems, the implications of this work aim to be far-reaching, contributing to better predictive models and a deeper understanding of the simplicity and complexity inherent in natural and artificial systems.

In this work, we extend the applications of simplicity bias to random dynamical systems, and in particular the (digitized) trajectories of the logistic map with additive noise \cite{doan2018hopf}. Our work is part of a broader research project into the relevance of simplicity bias and algorithmic probability to dynamical systems and time series forecasting  \cite{zenil2011algorithmic,dingle2024exploring,dingle2023note}, and indeed to mathematics and science more generally.  Furthermore, this study should be viewed in the same context as that presented in Ref.\ \cite{Allen2014}, where Martin-L\"of randomness was employed to examine Brownian motions, thereby demonstrating the utility of Algorithmic Information Theory (AIT) tools in analyzing random dynamical systems.


Most results in AIT, and the original algorithmic probability formulation, are given in terms of binary strings. While AIT is not restricted only to applications involving binary strings, applications involving binary strings are in a sense more natural and arguably easier to implement. Hence, in this study, we will digitize trajectories so as to study binary string trajectories (Cf.\ \cite{kanso2009logistic,dingle2024exploring,dingle2023note}). We begin by studying simplicity bias in the random logistic map and then investigate the effects of adding measurement noise also. The following section looks at complexity-probability graphs and simplicity bias in the noise-induced regime of the logistic map. Finally, we look at the logistic map in the context of the counter-intuitive claim of induction based on algorithmic probability, that sometimes more data, all of which follows the same trend, can result in \emph{less} confidence in our ability to extrapolate the trend.

\section{Background and problem set up}

\subsection{Background theory and pertinent results}

\subsubsection{AIT and Kolmogorov complexity}
At the intersection of computer science and information theory lies  \emph{algorithmic information theory} \cite{solomonoff1960preliminary,kolmogorov1965three,chaitin1975theory} (AIT). AIT concerns the information content of individual objects, such as binary strings, integers, and discrete geometries \cite{li2008introduction}. The central quantity of AIT is \emph{Kolmogorov complexity}, $K(x)$, which quantifies the shortest length of a program to produce some object $x$. Alternatively $K(x)$ can be thought of as the size in bits of the compressed version of $x$ (assuming a perfect compressor). An object $x$ which has a complex or random pattern is hard to compress, and so has a high Kolmogorov complexity, while an object $x$ containing simple repeating patterns like $x=01010\dots10101$ is highly compressible and so has a low Kolmogorov complexity value.

Because we will not be using or invoking detailed results from AIT, we will not go into details here giving formal definitions and proofs. There are many standard texts that the interested reader can refer to if needed, e.g., Refs.\ \cite{li2008introduction,calude2002information,gacs1988lecture,shen2022kolmogorov}.

\subsubsection{The coding theorem and algorithmic probability}
An important result in AIT is the coding theorem \cite{levin1974laws}, which establishes a connection between $K(x)$ and probability prediction estimates. It states that 
\begin{equation}
P(x) = 2^{-K(x)+O(1)}\label{eq:CD}
\end{equation}
where $P(x)$ is the probability that an output $x$ is generated by a generic computing device (technically a prefix optimal universal Turing machine) when fed with a random binary program. Thus, high complexity outputs have exponentially low probability, and simple outputs with small $K(x)$ values must have high probability. $P(x)$ is also known as the \emph{algorithmic probability} of $x$ (see Ref. \cite{hutter2007algorithmic} for an overview). AIT results are often difficult to apply directly in real-world contexts, due to a number of issues including the fact that $K(x)$ is formally uncomputable.

\subsubsection{The simplicity bias bound}

While the original theory regarding algorithmic probability is beautiful and powerful, as mentioned, it is not easy to directly apply the theory to practical real-world problems. However, several studies have tried to apply numerical approximations of the theory to concrete problems \cite{zenil2014correlation,zenil2011algorithmic,dingle2018input}.  Such applications have led to the observation of a phenomenon called \emph{simplicity bias} \cite{dingle2018input}. Simplicity bias has three levels of meaning, or we could say, precision: 
\begin{itemize}
    \item [(i)] A loose general inverse relation between probability and complexity
    \item [(ii)] A scatter plot of complexity vs.\ probability showing a roughly linear upper bound decay in log probability with increasing complexity 
    \item [(iii)] A complexity-probability plot as in (ii) but with the linear upper bound following the simplicity bias bound Eq.\ (\ref{eq:simplicity_bias}) (see below) with $a\approx 1$
\end{itemize}
If any of these are observed in a complexity-probability scatter plot, then it is considered as simplicity bias. The simplicity bias bound was presented in Ref.\ \cite{dingle2018input}, and has the form 
\begin{equation}
P(x)\leq 2^{-a\tilde{K}(x)-b}\label{eq:simplicity_bias}
\end{equation}
where $P(x)$ is the probability of observing output $x$ on random choice of inputs, and $\tilde{K}(x)$ is the estimated Kolmogorov complexity of the output $x$. The bound says that complex outputs from input-output maps have lower probabilities, while high-probability outputs must be simpler. The constants $a>0$ and $b$ can be fit with sampling and often even predicted without recourse to sampling \cite{dingle2018input}. Typically we assume that $a\approx 1$, which is the theoretical prediction. However, sometimes in practice, it can be hard to correctly estimate $a$ due to insufficient sampling. We will also assume throughout this work that $b=0$ in Eq.\ (\ref{eq:simplicity_bias}), which is a default assumption as argued and discussed in Ref.\ \cite{dingle2018input}. 

There is also a conditional version of the simplicity bias equation \cite{dingle2022predicting}. Because Eq.\ (\ref{eq:simplicity_bias}) is an upper bound, it allows for output patterns to fall below the bound. See Refs.\ \cite{dingle2020generic,alaskandarani2023low} for more on these types of patterns.

Simplicity bias and the related bound have been found to apply in many systems including RNA shapes \cite{dingle2018input,dingle2023multiclass}, protein shapes \cite{johnston2022symmetry,dingle2022predicting}, models of plant growth \cite{dingle2018input}, ordinary differential equation solution profiles \cite{dingle2018input}, finite state transducers \cite{dingle2020generic}, natural time series \cite{dingle2023note}, and others. In these systems, the probability of different output shapes, upon a random sampling of inputs, was non-trivially bounded and thereby predicted by Eq.\ (\ref{eq:simplicity_bias}). 

The main motivations for studying simplicity bias are (a) uncovering a `universal' property of input-output maps; (b) identifying a basic and general mechanism for the appearance of simplicity and symmetry in nature; and (c) exploring a fundamentally different way to predict the probability of output shapes as compared to standard methods like frequency sampling. Simplicity bias means that some predictions about output probabilities can be made directly by estimating the complexity of outputs, largely independently of the underlying system mechanisms and without having to make frequency counts.

\subsection{Digitised map trajectories}
AIT results, including algorithmic probability, are typically given in terms of random inputs or `programs' which are computed to produce output strings or patterns. Further, these results of AIT are normally given in terms of binary strings. While dynamical systems may not appear to fit this input-output framework, they can be viewed as input-output functions if the initial value and parameters are considered input `programs' which are then computed to generate `output' dynamical system trajectories. As above, because output pattern complexities and probabilities are more straightforwardly calculated if outputs are binary strings, we will digitize the real-valued trajectories into binary strings.

In Ref.\ \cite{dingle2024exploring} we explored simplicity bias in the standard deterministic logistic map with equation
\begin{equation}
x_{k+1}=\mu x_k(1-x_k) \label{eq:logisticmap}
\end{equation}
where the inputs are the pair of values $x_0\in$(0.0, 1.0) and $\mu\in$ (0.0, 4.0], and $k=1,2,3,\dots,n$. Here, we will extend the analysis to the logistic map with added noise, as studied by Sato et al.\ \cite{sato2018dynamical}. This random logistic map has the form
\begin{equation}
x_{k+1}=\mu x_k(1-x_k) +\omega_k \label{eq:randomlogisticmap}
\end{equation}
where $\omega_k$ is an i.i.d.\ random number sampled uniformly from $[-\epsilon,\epsilon]$, so that $\epsilon$ determines the magnitude of the noise. 

The random logistic map can be viewed as an input-output map because the values of $\mu$ and $x_0$ are like `inputs' which are computed (in addition to the noise) to produce a binary sequence trajectory `outputs'. In more detail, the map is first iterated to obtain a sequence of $n$ real values $x_1,x_2,x_3,\dots,x_n$ in [0,1]. Similar to the field of symbolic dynamics \cite{lind1995introduction}, the real-valued trajectory is digitized to become a binary string output sequence $x$ by applying a threshold, writing 1 if $x_k\geq0.5$ and 0 if $x_k<0.5$ \cite{kanso2009logistic,kaspar1987easily,dingle2024exploring}. Hence a binary sequence output $x$ of $n$ bits is generated for each input pair $\mu$ and $x_0$, and random realization of $\omega_k$ for $k=1,2,3,\dots,n$. 

For example, some choice of $\mu$, $x_0$, and $\epsilon$ after iterating Eq.\ (\ref{eq:randomlogisticmap}) with $k=1,2,...,25$ might yield the real-valued trajectory 
\[x_1, x_2,\dots, x_{25} = 0.12, 0.47, 0.66, \dots , 0.21, 0.05, 0.78, 0.97\]
which after digitization becomes the binary string
\begin{equation}
  x= 001\dots0011  
\end{equation}

Following Ref.\ \cite{dingle2024exploring}, we will use $n=25$, but see the same reference for the impact of larger or smaller $n$. In brief, $n$ should be large enough to yield many different patterns and complexity values, but small enough so that decent frequency and probability estimates can be made for up to $2^n$ different strings without the need for excessive sampling. 

Simulations will proceed by fixing some choice of $\mu$ and $\epsilon$, and then randomly sampling $x_0\in (0.0, 1.0)$, and iterating for $n=25$ steps. Each sample will yield a binary string $x\in\{0,1\}^n$, i.e., a binary string of length $n$. By counting the relative frequency of strings $x$, we can find $P(x)$, the probability of each string $x$. The estimated complexity $\tilde{K}(x)$ of each string can be calculated using a compression-based metric, based on the Lempel-Ziv 1976 complexity measure \cite{lempel1976complexity} (see Appendix for more details on the complexity measure). Using the values of $P(x)$ and $\tilde{K}(x)$ for each binary string, we make a scatter plot of these quantities and examine the connection (if any) between complexity and probability.

\section{Results}

\subsection{Simplicity bias in the random logistic map}

In this study, we are primarily interested in  $\mu$-values for which the trajectories of the deterministic map converge quite quickly to fairly trivial fixed points or periods.  If instead $\mu\approx4.0$ then it is known that there is no simplicity bias in trajectories \cite{dingle2024exploring}, so adding noise will only serve to increase the complexity and variability of the trajectories, making any kind of bias and therefore simplicity bias, not appear. Note that we use the word `bias' to describe a strongly non-uniform (low entropy) probability distribution, such that certain binary string outputs are much more likely the others.

It is known that if $\mu\in[0.0,1.0]$, then the trajectories of the deterministic logistic map tend to 0 \cite{hasselblatt2003first}.  Because we are adding noise to the trajectories, the trajectories will be prevented from becoming arbitrarily close to zero. If the noise is small enough, then $x_k\approx0 <0.5$ so that the resulting digitized binary string (after truncating at 0.5) will be $x=...0000$. If the noise is larger, then it may be that $x_k>0.5$ for some $k$, so that more varied patterns including 1s can emerge.
For $\mu\in(1.0,3.0]$  the trajectories of the deterministic logistic map tend to a fixed point with value $1-(1/\mu)$  \cite{berger2001chaos}. If $\epsilon=0.0$, then for any initial value of $x_0$ the trajectories will tend to be either $x=\dots000$ or $x=\dots111$ depending on $\mu$, which are trivial patterns. Not only are they trivial, they are not varied enough to have different complexity and probability values, and so the corresponding complexity-probability plot is trivial. Therefore, simplicity bias cannot emerge in this scenario, at least if we ignore the initial transient dynamics (but see also Ref.\ \cite{dingle2024exploring} for more on this). However, if we add noise then different and non-trivial patterns can emerge. 

\begin{figure*}[htp]
\begin{center}
\subfigure[]{\includegraphics[height=5.5cm,width=5.5cm]{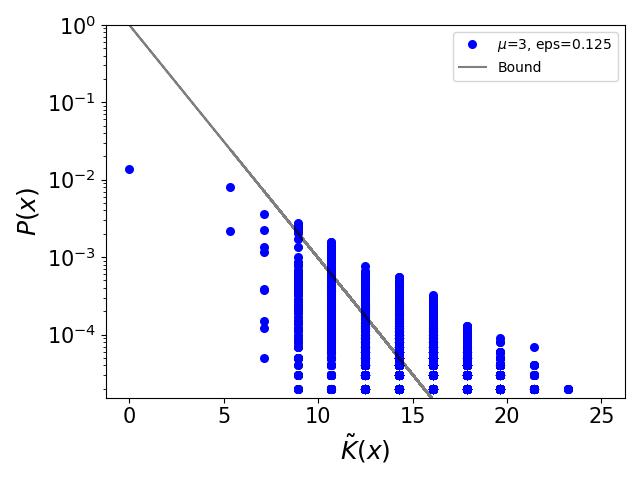}}
\subfigure[]{\includegraphics[height=5.5cm,width=5.5cm]{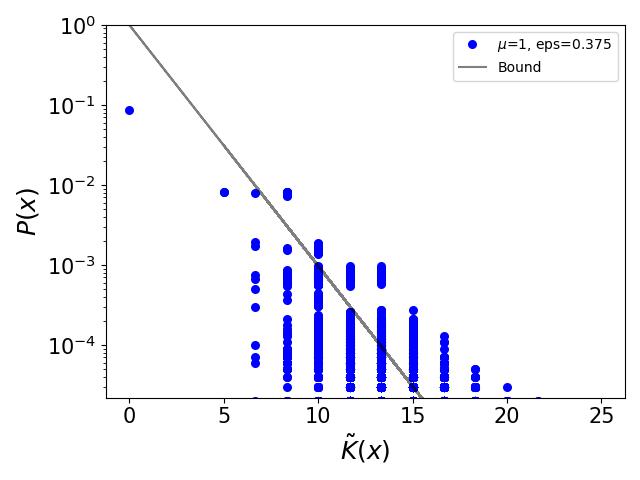}}
\subfigure[]{\includegraphics[height=5.5cm,width=5.5cm]{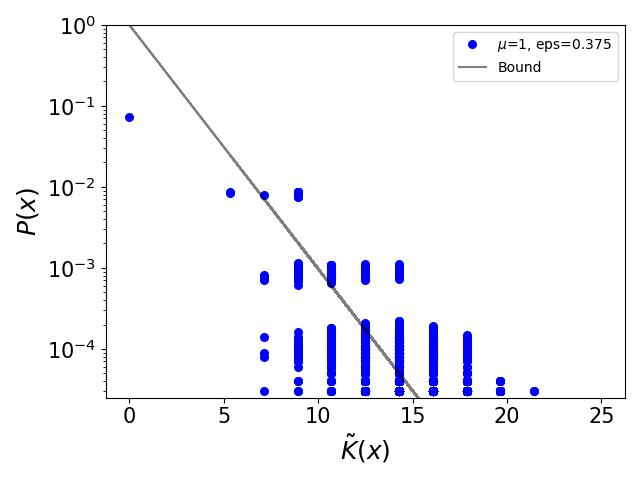}}
\end{center}
\caption{Simplicity bias in the digitized random logistic map. Each blue data point corresponds to a different binary digitized trajectory $x$ of length 25 bits. The black line is the upper bound of Eq.\ (\ref{eq:simplicity_bias}) with $a=1$ as a default prediction. (a) $\mu=3.0$ and $\epsilon=0.125$ including transient dynamics; (b) $\mu=1.0$ and $\epsilon=0.375$ including transient dynamics; and (c)  $\mu=1.0$ and $\epsilon=0.375$ excluding transient dynamics.
}
\label{fig:simplicity_bias_RDS}
\end{figure*}

The parameter values used in the numerical experiments are $\mu=$ 0.0, 1.0, 2.0, and 3.0 in combination with $\epsilon=$ 0.125, 0.25, 0.375, and 0.5. For many of these parameter combinations, we can expect no simplicity bias in the binary trajectories, due to the following: If the noise $\epsilon$ is large, then the trajectory is essentially a random erratic pattern across the interval [0,1] with little correlation between adjacent bits, yielding a roughly uniform distribution over strings. Therefore, there is unlikely to be bias for some specific patterns, and consequently no bias for simple patterns. On the other hand, if $\epsilon$ is very small, then the `kicks' imposed by the noise will not be large enough to prevent the trajectories from converging. As examples, if $\mu=1.5$ and $\epsilon\approx 0$, then the trajectory values $x_k$ will fluctuate around the value $1-(1/\mu)=0.33<0.5$, yielding a binary trajectory $x=...0000$. If $\mu=2.0$, then the fixed point would be $1-(1/\mu)=0.5$, so even small fluctuations around this value would yield almost random binary strings $x$. From this reasoning, we can conclude that to observe simplicity bias in the random logistic map requires having a balance of noise that is not too large as to overly randomize the trajectories, and not too small as to allow the trajectories to converge to a fixed point with only modest fluctuations around $1-(1/\mu)$.

Turning to numerical experiments, we make complexity-probability plots for different values of $\mu$ and noise magnitude values $\epsilon$. We separately analyze the cases of including initial transient dynamics before the trajectories settle, and ignoring the transient dynamics by excluding the first 50 iterations of the map. Ignoring the initial iterations is common practice in the mathematical physics community \cite{berger2001chaos}. However, from the machine learning perspective, the initial transient dynamics where a larger fraction of the space is sampled is more important for statistical learning. So we will consider both regimes.

Figure \ref{fig:simplicity_bias_RDS} shows some examples of simplicity bias. The different binary string outputs that appeared in sampling are shown as blue dots. The black line bound of Eq.\ (\ref{eq:simplicity_bias}) is also depicted with $a=1$ (the default prediction).
Figure \ref{fig:simplicity_bias_RDS}(a) shows $\mu=3.0$ and $\epsilon=0.125$, and the initial dynamics are included. There is simplicity bias because we see that some low complexity patterns have high probability. 
Further, we see a roughly linear upper bound decay in the probabilities, as complexity $\tilde{K}(x)$ increases. However, the gradient appears to be less steep than the $a=1.0$ default prediction suggests.  The slope of the upper bound is -0.27. The reason that the slope is not accurate may be due to undersampling because estimating the slope $a$ assumes that nearly all of the possible strings have appeared in the sample, which may be overly coarse. Another reason may be due to the introduction of random noise (see next section for more on this). Despite these considerations, it is still clear that simplicity bias is observed. Without the additive noise, the trajectory would have converged to $1-1/\mu=0.67$ and shown no simplicity bias.

In Figure \ref{fig:simplicity_bias_RDS}(b) we see data for $\mu=1.0$ and $\epsilon=0.375$, and the initial dynamics are included. Again there is clear simplicity bias. Without the additive noise, the trajectory would have converged to $1-1/\mu=0.0$ and shown no simplicity bias.
In addition to the inverse relation between probability and complexity, even the gradient of the decay of the upper bound is not strongly divergent from the black line default prediction. The slope of the upper bound is -0.32. In Figure \ref{fig:simplicity_bias_RDS}(c) we see data again for $\mu=1.0$ and $\epsilon=0.375$, but this time after ignoring the first 50 iterations (to exclude the transient dynamics).  Here we see simplicity bias in the form of an inverse relation between probability and complexity. Having said that, the simplicity bias in panel (c) is not as pronounced as in panel (b). The slope of the upper bound is -0.37.

For each complexity value, we see outputs with a range of probabilities. That is, there are some patterns that are simple, yet have very low probability values. These low complexity, low probability outputs have been explored in detail earlier \cite{dingle2018input,alaskandarani2023low}. However, it is known that these low complexity, low probability outputs collectively do not account for much of the probability mass \cite{dingle2020generic}. 

In conclusion, we see that for some parameter values, simplicity bias is observed in the random logistic map trajectories, even while the deterministic counterparts with the same $\mu$ values would have converged to fixed points and not shown simplicity bias.

\begin{figure*}[htp]
\begin{center}
\subfigure[]{\includegraphics[height=5.5cm,width=5.5cm]{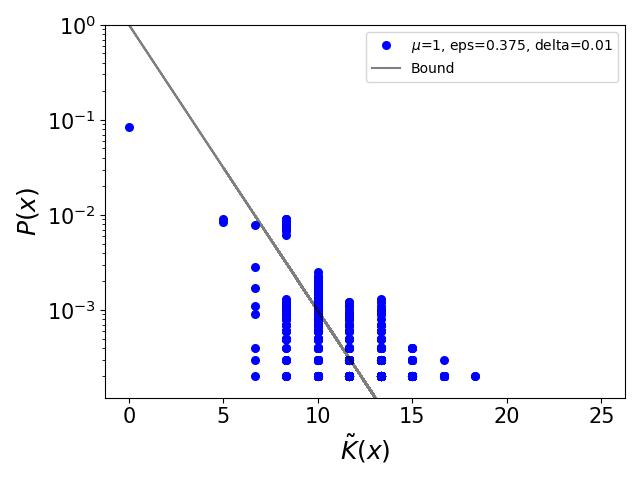}}
\subfigure[]{\includegraphics[height=5.5cm,width=5.5cm]{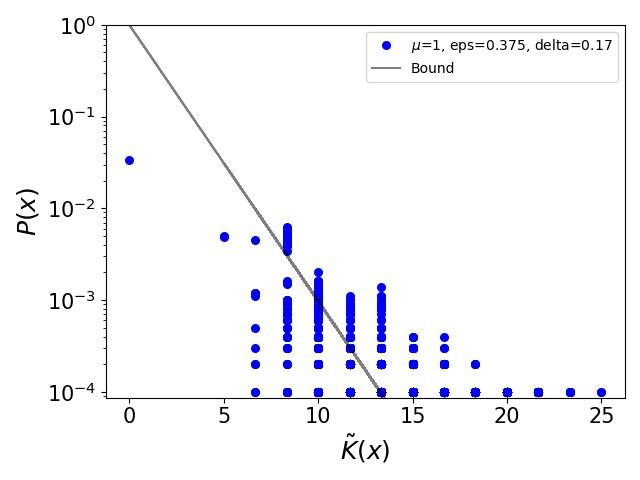}}
\subfigure[]{\includegraphics[height=5.5cm,width=5.5cm]{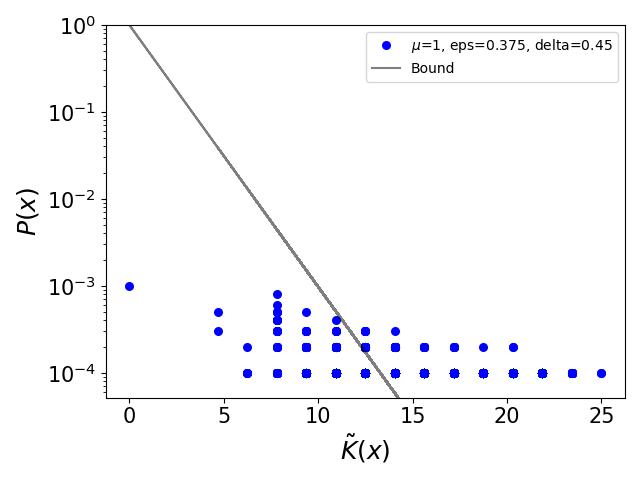}}
\end{center}
\caption{Simplicity bias with measurement noise. In (a), a very small amount of measurement noise ($\delta=0.01$) is added, and simplicity bias is clear. In (b) a larger amount of noise is added ($\delta=0.17$), and simplicity bias remains. In (c) a larger value of noise is added ($\delta=0.45$), and the slope becomes much less steep, thus weakening the probability-complexity connection.}
\label{fig:noisymeasurements}
\end{figure*}

\subsection{Adding noise to the strings}

There are different ways in which noise can be included in the modeling of the random dynamical system. Above, we included additive noise in the dynamics themselves, to model the scenario in which the dynamics themselves are not deterministic but perturbed stochastically. In this section, we now consider another type of noise,  measurement noise, which models the scenario that when measuring the trajectory, we do so inaccurately.  

It is important to study if simplicity bias is maintained under different types of perturbation, because if it does not, then it may not be a physically relevant body of theory. This is related to arguments supporting the study of random dynamical systems, over deterministic systems, because some results from chaos theory disappear in the presence of even small noise, suggesting that they are not relevant to the study of physical dynamical systems (which are inherently stochastic, typically).

We introduce measurement noise as follows: If $x_1x_2x_3\dots x_n$ is a  trajectory of real-values, then we replace  each $x_k$ with $x_k+\delta_k$, where $\delta_k$ is some small uniform noise in $[ -\delta,\delta]$. After adding the noise, we use the 0.5 threshold as above and digitize the noisy trajectory. We can then, as above, make a complexity-probability plot and look for simplicity bias.

In Figure \ref{fig:noisymeasurements} we see the effects of adding measurement noise. For illustration purposes,  we use the same values of $\mu$ and $\epsilon$ as in Figure \ref{fig:simplicity_bias_RDS}, and increasing values of $\delta$. In Figure \ref{fig:noisymeasurements}(a), a very small noise level of $\delta=0.01$ is used, and hence the simplicity bias is still clearly observed. In (b), a larger noise level of $\delta=0.17$ is used, and we see that the simplicity bias still remains, but the slope of decay in log probability with increasing complexity is slightly less steep as compared to (a).
Finally, in (c), a larger noise level of $\delta=0.45$ is used, and the bias and simplicity bias has almost completely disappeared with only a modest slope decay in log probability. The slopes of the upper bounds are  -0.12, -0.10, and  -0.04 respectively.
In sum, the simplicity bias gradually disappears with increasing $\delta$ via the slope gradually becoming more horizontal. 

It is noteworthy that in some earlier studies of simplicity bias, the slope of the decay in log probability was observed to be considerably less steep than the upper bound of Eq. (\ref{eq:simplicity_bias}) but it was not stated why this was the case \cite{dingle2023note,dingle2020generic}. The results here suggest a simple explanation for this observation, i.e.,  noise (in some form) corrupts the simplicity bias, adjusting the slope.

\subsection{Noise-induced chaos around $\mu=3.83$}

For a large fraction of values of $\mu\geq 3.56994567...\approx 3.57$ (\texttt{https://oeis.org/A098587}), the trajectories of the deterministic logistic map are chaotic \cite{hasselblatt2003first}. However, there are also islands of stability for some $\mu\in(3.56994567..., 4.0]$, in which the dynamics are simple and periodic. One example is when $\mu=3.83$, which has an attracting period-3 trajectory in the deterministic case. It has been known for many years, however, that with a small amount of noise added to the random logistic map, noise-induced chaos can appear \cite{mayer1981influence,crutchfield1982fluctuations}. That is, with $\mu=3.83$ it is known that $\epsilon \approx 0.00146$ is the critical noise amplitude at which the chaotic attractor appears. At this value, the Lyapunov exponent becomes positive. 

We will study this noise-induced chaos in terms of simplicity bias. Above, we saw that simple fixed point dynamics combined with additive noise can lead to simplicity bias in the trajectories. This suggests that the simple period-3 trajectory when $\mu=3.83$ may show simplicity bias when combined with additive noise of $\epsilon=0.00146$. In contrast, for the chaotic regime, the logistic map does not exhibit bias, nor simplicity bias \cite{dingle2024exploring}, because no particular patterns are much more likely (and therefore no simple patterns, either). Therefore, we will also examine $\epsilon=0.00146$ combined with $\mu=3.82$ and $\mu=3.84$ which potentially will not show simplicity bias, assuming that the dynamics are chaotic for these values.

\begin{figure*}[htp]
\begin{center}
\subfigure[]{\includegraphics[height=5.5cm,width=5.5cm]{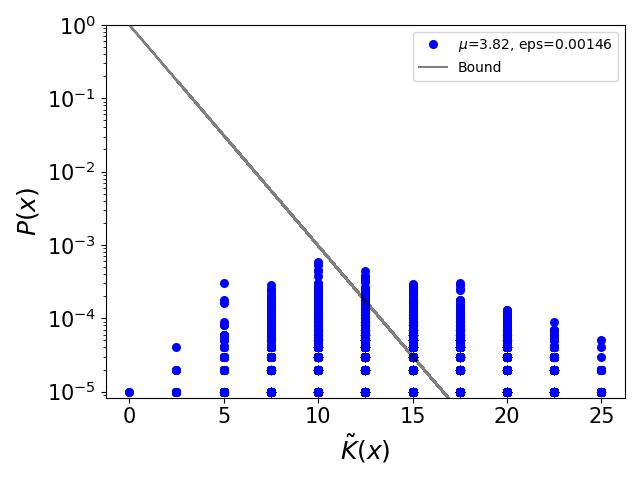}}
\subfigure[]{\includegraphics[height=5.5cm,width=5.5cm]{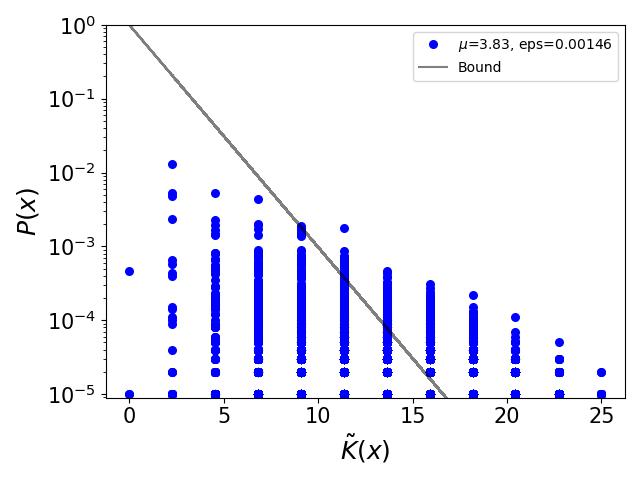}}
\subfigure[]{\includegraphics[height=5.5cm,width=5.5cm]{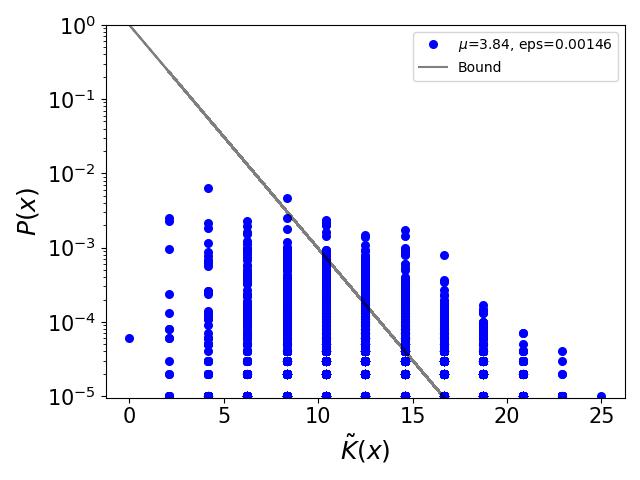}}
\end{center}
\caption{Complexity-probability plots for $\mu \approx 3.83$. (a) With $\mu=3.82$ there is no simplicity bias; (b) with $\mu=3.83$ there is clear simplicity bias with a roughly linear upper bound decay; and (c) with $\mu=3.84$ there is still simplicity bias, but the linear upper bound decay is more noisy than with $\mu=3.83$.  }
\label{fig:mu383_RDS}
\end{figure*}

We perform numerical experiments as above, digitizing the trajectories. Figure \ref{fig:mu383_RDS}(a) shows the complexity-probability plot for $\mu=3.82$, and there is no evidence of simplicity bias, as expected. There is some bias in the distribution, as inferred from the fact that some binary strings have probability $\approx10^{-3}$, while some others have probability $\approx10^{-5}$. However, the connection to complexity is very weak. In (b) however, with $\mu=3.83$ there is clear evidence of simplicity bias, with a linear upper-bound decay in log probability with increasing complexity. The slope of the decay is not the same as the black line bound, but somewhat less steep. In Figure \ref{fig:mu383_RDS}(c), interestingly with $\mu=3.84$ there is still simplicity bias, but not quite as clear as in (b) as evidenced by the less clearly linear upper bound.

While there are already other ways to detect and measure noise-induced chaos \cite{mayer1981influence}, this perspective of simplicity bias also provides another perspective.

\subsection{More data does not always give more confidence in predictions}

In the preceding sections, we looked at how simplicity bias, which is a form of algorithmic probability, can be used to predict or bound the probability of patterns in time series from the logistic map. Another application of algorithmic probability, and in fact one of the original motivations for its development, is for inference or induction of series. Solomonoff \cite{solomonoff1960preliminary,solomonoff1964formal} was interested in questions like, given a (finite) binary series $x$ representing a historical series, what is the best prediction for the next bit in the series? More quantitatively, what is the probability that the next bit is a zero, $P(0|x)$, and what is the probability the next bit is a 1, $P(1|x)$? If the underlying process generating the series is known, e.g., it is a Bernoulli process, then standard statistical theory can give methods to estimate the probabilities of $P(0|x)$ and $P(1|x)$. In many practical situations, however, the underlying process is not known, which motivates the development of forecasting methods which do not depend on knowing the process. So, is there a general way to predict time series, which does not require (strong) assumptions or knowledge about the process that made the series? Solomonoff induction \cite{li2008introduction} is the method proposed by AIT, which is essentially prediction based on algorithmic probability. 

Hutter \cite{hutter2004universal} has also argued that algorithmic probability should be central to artificial intelligence and the theory of machine learning. The main problem with algorithmic probability, and hence Solomonoff induction, is that it is formally uncomputable \cite{li2008introduction}, so cannot be applied directly in practice (or at least not straightforwardly). Indeed, some have argued that this makes the theory almost useless \cite{neth2023dilemma}. However, developing the theory is useful because some aspects can shed light on real-world problems, and some of the theory can be approximated and thereby applied. 

Here we will consider one rather surprising implication of Solomonoff induction: sometimes with \emph{more} data --- even if the data points all conform to one trend --- we should be \emph{less} confident in our ability to predict the series (see Li and Vitanyi \cite{li2008introduction} for a mention of this result, and more in Hutter \cite{hutter2007universal}). We will see how this result can apply to the logistic map. To our knowledge, the logistic map example presented below is the first concrete example illustrating this implication.

\subsubsection{Theory}
Suppose we observe some binary series, which happens to be just 1 repeated $n$ times, i.e., $1^n=$ 111...111. What should be our prediction for the next bit? And how confident should we be of our prediction? Laplace famously argued that if there are $k$ 1s out of $n$ binary observations, then we should predict that the next bit is 1 with probability 
\begin{equation}
    P(1|111...111) = \frac{k+1}{n+2} = \frac{n+1}{n+2} = 1-\frac{1}{n+2}\approx 1-1/n
\end{equation}
where we have set $k=n$ because all observations are `1'. According to Laplace then, if we observe more data (larger $n$) all of which conforms to the same trend (all 1s), then we should have greater confidence that the next bit will also be 1 (because $1-1/n$ is closer to one). This reasoning also accords with common-sense statistics.
However, using a general form of inference based on Solomonoff induction and algorithmic probability, we do not always arrive at the same conclusion. 

Solomonoff induction (or prediction) is based on imagining that a computer has been fed a random input program, and it produces an infinite binary string output. We observe the first $n$ bits of the output as a binary string $x$, and we try to predict the probability that the next bit is a 0 or a 1. Technically this computer should be a monotone universal Turing machine, but we will not explore the details of this because they are not essential to understanding our example; see \cite{hutter2007algorithmic,hutter2007universal,li2008introduction} for more details. We will write $M(x)$ for the probability that the infinite string begins with $x$, and write the conditional probability as
\begin{equation}
    M(x_{n+1}|x_1x_2x_3\dots x_n) = \frac{M(x_1 x_2 x_3\dots x_n x_{n+1})}{M(x_1 x_2 x_3\dots x_n)}
\end{equation}
Essentially, $M(x_{n+1}|x_1x_2x_3\dots x_n)$ gives the conditional probability prediction based on compressing the sequence $x_1 x_2 x_3\dots x_n x_{n+1}$, and assigns the predicted probabilities of $M(0|x_1x_2x_3\dots x_n)$ and $M(1|x_1x_2x_3\dots x_n)$ based on how well the following bit (0 or 1)  fits the preceding sequence, as measured by how well the following bit (0 or 1) compresses the sequence.

Using algorithmic probability now as a method for induction or prediction, the probability that the next bit in the series after observing $x=111\dots 111$ is 1 is given by \cite{hutter2007universal}
\begin{equation}
    M(1|111...111)=1 - 2^{-K(n)+O(1)}
\end{equation}
For most $n$, or typical $n$, it is known that the complexity of an integer $n$ is given by $K(n)\approx \log_2(n)$ to first order \cite{li2008introduction}, so that 
\begin{equation}
    M(1|111...111)= 1 - 2^{-K(n)+O(1)}\sim 1- 2^{-\log_2(n)} =  1 - 1/n
\end{equation}
roughly agreeing with Laplace. 

A difference in predictions between the Laplacian and algorithmic probability approaches can arise when certain values of $n$ are encountered, which are `special', and therefore (algorithmically) `simple'. Such values of $n$ are those that can be described succinctly and unambiguously. For example, if $n=2^{2^{2^{2^2}}}\approx 2\times 10^{19728}$ is written out in full it has nearly 20,000 digits (in base 10), while at the same time we can describe it very succinctly in just a few characters using a stacked exponential. Therefore we say that $n$ is simple, and the Kolmogorov complexity of this value $n$ is low. As another example of `special' or `simple' integers, consider the factorial $n=(8!)!\approx 3 \times 10^{168186}$. This integer has nearly 170,000 digits if written out explicitly in full, yet can be described exactly in just a few characters. There are many `special' and algorithmically simple numbers, but nonetheless, they are rare in the sense that a vanishingly small fraction of integers are special or simple \cite{li2008introduction}.

For such `special' or `simple' numbers $n$, the Kolmogorov complexity is small. It follows that for a simple $n$ which has $K(n)\ll \log_2(n)$,  the predicted probability for the next bit is
\begin{equation}
    M(1|111...111)\approx 1 - 2^{-K(n)+O(1)}< 1-1/n
\end{equation}
which can be substantially different from Laplace's estimate for large enough $n$. This is not to say that algorithmic probability would predict a 0 in place of a 1 as the next bit, but instead, it means that the prediction that the next bit is 1 --- for this $n$ --- is less confident (further from 1) than it would be for a typical $n$. This is a kind of quantification of 
the intuition that typically we do not expect there to be a change in a series after a certain pattern has been followed for a long time (e.g., 1s to change to 0s), but on `special occasions' when $n$ is somehow special, then it is more likely that something will `happen' (i.e., a change). 
 
Rather surprisingly, what this means also is that for some few special values of $n$, more data does not necessarily mean higher confidence in our predictions: 
consider $n$ is some large random integer such that $K(n)\approx \log_2(n)$, while $n'$ is some other simple integer such that $n'>n$. It follows that
\begin{equation}
    K(n')\ll \log_2(n) < \log_2(n') \Rightarrow M(1|1^{n})>M(1|1^{n'})
\end{equation}
Thus, even though we have observed $n' - n$ more 1s repeated, we are \emph{less} confident in our prediction that the next bit is 1, because $M(1|1^{n})$ is closer to 1 than $M(1|1^{n'})$. 

\subsubsection{Example}

We will now consider how this somewhat counterintuitive result from algorithmic probability theory applies to the logistic map. 
We will construct an example of when the trend of just 0s changes after a long time and becomes a string of 1s at an algorithmically simple value of $n$.

Suppose we choose $\mu=2.5$, $\epsilon=0.0$, and $(x_0)^{-1}=2^{2^{2^{2^2}}}$ so that $x_0\approx 10^{-19728}$, then this series begins with $x_0<0.5$. Because $\mu=2.5$, the fixed point which the trajectory approaches is $1-1/\mu=1-1/2.5=0.6>0.5$. Therefore, eventually, the digitised series will change from 0s to 1s. Call $n^*$ the integer at which the series $0000...011111$ transitions from 0s to 1s.

Because $x_0\approx 0$ and $\epsilon=0.0$, we have $(1-x_k)\approx 1$ for $k\approx 1$ so that the logistic equation Eq.\ (\ref{eq:logisticmap}) simplifies to $x_{k+1}\approx 2.5 x_k$, or $x_{k+1}\approx 2.5^k x_0$. This implies that when $x_k\approx 0$ the $x_k$ terms grow exponentially fast initially (but slows as $x_k$ gets larger). 
Even with exponential growth, $n^*$ must be a large number because $x_0$ is so small, indeed because $ 10^{19728} \approx 2.5^{49575}$, the number of iterations $n^*$ required to reach the 0.5 threshold is at least of the order of 50,000.  

So, having observed a series of $n^*$ 0s, according to Laplace's induction method, the probability that the next bit is also 0 will be roughly estimated as 
\begin{equation}
    P(0|000...000) = \frac{k+1}{n+2} = \frac{n^*+1}{n^*+2} = 1 - 1/n^* > 1- 1/50000 = 0.99998
\end{equation}
so that $P(1|000...000) < 0.00002$. Therefore, the sudden change in the logistic map from 0s to 1s at $n^*$ would come as rather a `surprise' for Laplace. 

In contrast, using algorithmic probability we have the following reasoning: Because $x_0$ is an algorithmically simple (small) number, and $\mu$ is a simple number also, and the logistic equation Eq.\ (\ref{eq:logisticmap}) can be written in only a few characters also, and because the value of $n^*$ can be derived via a simple function of these values (e.g., by iterating the logistic map to find the value of $n^*$), then this implies that $n^*$ is a special number and $K(n^*)$ is small\footnote{Technically, the value of the complexity $K(n^*)$ will depend on the specific computer of language used, but we ignore this for this example. Later we give an example which does not depend on these.}. Therefore we have 
\begin{equation}
  K(n^*)< \log_2(n^*) \Rightarrow  M(0|000...000)= 1 - 2^{-K(n^*)+O(1)}< 1-1/n^*
\end{equation}
so that $M(1|000...000) > 1/n^*$, so that the change from 0s to 1s is less of a `surprise' if using the algorithmic probability approach. It follows also that we are less confident about predicting the next bit in the sequence $1^{n^*}=111\dots111$, as compared to predicting the next bit in the sequence $111\dots111$ where the length of the sequence is some algorithmically random but slightly smaller value than $n^*$.

Note that the algorithmic probability estimate does not give an exact figure for the probability of changing to 1s after the 0s (which would depend on the specific computer or language used), but it does reveal that the probability prediction can be significantly different when $n$ is simple compared to typical values of $n$. See the Appendix for an example that does not depend on the specific computer or language used.

\section{Discussion}

Arguments inspired by algorithmic information theory (AIT) have been used to predict the occurrence of \emph{simplicity bias} in many real-world input-output maps, where complex patterns have exponentially low probabilities, and high probability patterns are simple \cite{dingle2018input}. 
This phenomenon has been observed in a very wide range of systems, including RNA shapes \cite{dingle2018input,dingle2023multiclass}, protein shapes \cite{johnston2022symmetry,dingle2022predicting}, models of plant growth \cite{dingle2018input}, ordinary differential equation solution profiles \cite{dingle2018input}, finite state transducer \cite{dingle2020generic}, natural time series \cite{dingle2023note}, deep neural networks \cite{valle2018deep}, 1D dynamical systems \cite{dingle2024exploring}, and others. 
Relatedly, Zenil and Delahaye \cite{zenil2011algorithmic} numerically studied the frequencies by which very short ($\sim$5 bits) binary string motifs appear in nature, namely in hard drives, DNA fragments,  and real-world image data. They found a correlation between these short binary string motifs, and the frequencies with which the same motifs appeared from generic computation devices (e.g., cellular automata). Their investigation showed that some simpler binary strings occurred with higher probability, hence a form of simplicity bias in natural data.

In this work, we have numerically investigated the presence of simplicity bias in digitized trajectories of the random logistic map. This current work sits in the context of a wider research program of investigating the `universal' presence of simplicity bias in input-output maps, which can aid in an understanding of the origins of symmetry and simplicity in nature, and also serve the practical aim of aiding \emph{a priori} probability predictions, based on directly measuring the complexity of patterns (instead of e.g., sampling).

Our main conclusions are that (i) we observe simplicity bias in the random logistic map for some restricted set of parameter values; (ii) adding measurement noise to trajectories gradually erases simplicity bias by flattening the slope in complexity-probability plots; (iii) the regime of noise-induced chaos when $\mu=3.83$ shows-up in complexity-probability plots, and (iv) algorithmic probability-based induction shows how sometimes in the logistic map trajectories, counterintuitively more data does not improve our confidence in extrapolating the trend.

The presence of a bias towards simplicity in this example of the logistic map with additive noise could be explained in a mechanistic fashion, and indeed many of the other examples of simplicity bias may admit mechanistic explanations based on the details of the specific way inputs are assigned to outputs. Despite this, we propose that it is beneficial to explain the observed simplicity bias from another perspective entirely, that of information theory and the simplicity bias bound Eq.\ (\ref{eq:simplicity_bias}). This latter perspective forms a unifying theory to unite the disparate observations in different maps. These different types of explanations are complementary.

It is noteworthy that digitized logistic map trajectories have been studied in terms of Lempel-Ziv complexity by  Kaspar and Schuster \cite{kaspar1987easily}, and in this sense is related to our work. Despite this, their work is fundamentally different from ours because it is not concerned with simplicity bias or in estimating the probability of different outputs. Also, Zenil et al.\ \cite{zenil2019algorithmic} have studied discrete dynamical systems of cellular automata from the perspective of algorithmic information theory, including reconstructing the space-time evolution of the systems. However, their work is different several respects, including that they did not look at the (random) logistic map in terms of simplicity bias.

The theory of algorithmic probability was developed several decades ago \cite{solomonoff1960preliminary,solomonoff1964formal,solomonoff1978complexity,levin1974laws}, and is important and potentially widely applicable in science and mathematics \cite{li2008introduction,hutter2004universal,vitanyi2013similarity,hutter2007algorithmic,zenil2019coding}. Applications of the field have been held back, partly at least, by the difficulty of applying the theory due to the incomputability of important quantities. However, it has been argued that algorithmic probability predictions should ``serve as ‘gold standards’ that practitioners should aim at ... [but which] have to be (crudely) approximated in practice'' \cite{hutter2007universal}. The simplicity bias bound, including the results here, can be seen in this light. Further, many other researchers have built on and applied the compression-learning-prediction link \cite{adriaans2007learning}. Indeed, very recently it has been argued that even the currently highly influential ``large language models [e.g., ChatGPT] are powerful general-purpose predictors and that the compression viewpoint provides novel insights'' \cite{deletang2023language}. It remains to be seen in the near future whether this intriguing compression approach provides valuable outcomes for machine learning.

In future work, we suggest continuing the application of (approximations to) algorithmic probability to develop novel approaches to probability predictions and advance statistical and machine learning methods.

\section{Conclusion}

In this study, we have demonstrated the presence of simplicity bias within the random logistic map, affirming its role across a range of dynamic systems. We observed that simplicity bias persists within specific parameter regimes but is gradually diminished by measurement noise, particularly within noise-induced chaos regimes. These findings not only reinforce the intricacies of dynamical systems but also emphasize the counterintuitive nature of algorithmic probability-based predictions.

Our research underscores the value of interpreting simplicity bias through an information-theoretic lens, offering a unifying framework that transcends specific system mechanics. This perspective aligns with broader efforts to employ algorithmic probability as a theoretical standard in understanding complex systems, despite its practical challenges.

As we look forward, the application of algorithmic probability and related concepts remain a promising frontier for advancing prediction and analysis methods in various fields. By continuing to bridge theory with practical applications, we aim to deepen our understanding of the principles that govern simplicity and complexity in the natural world.



\vspace{0.5cm}
\noindent
{\bf Acknowledgments:} BH thanks Prof. Jeroen Lamb (Imperial College London) for useful discussions about random dynamical systems that inspired the work in this paper. BH  acknowledges support from the Jet Propulsion Laboratory, California Institute of Technology, under a contract with the National Aeronautics and Space Administration and from Beyond Limits (Learning Optimal Models) through CAST (The Caltech Center for Autonomous Systems and Technologies).

KD thanks Muhammad Alaskandarani for useful discussions and work on the early parts of this work. This work has been partially supported by the Gulf University for Science and Technology, including by project code: ISG Case 9.

\bibliographystyle{unsrt}
\bibliography{SB_RDS_Refs, reference, missing_dyn} 

\def\cprime{$'$}
\begin{thebibliography}{10}

\bibitem{5706920}
Jake Bouvrie and Boumediene Hamzi.
\newblock Balanced reduction of nonlinear control systems in reproducing kernel
  hilbert space.
\newblock In {\em 2010 48th Annual Allerton Conference on Communication,
  Control, and Computing (Allerton)}, pages 294--301, 2010.

\bibitem{}
England and Wales~Court of~Appeal (Civil~Division).
\newblock Nulty \& {O}rs v.\ {M}ilton {K}eynes {B}orough {C}ouncil, 2013.
\newblock [2013] EWCA Civ 15, Case No. A1/2012/0459.
  \url{http://www.bailii.org/ew/cases/EWCA/Civ/2013/15.html}.

\bibitem{yk4}
R.~González-García, R.~Rico-Martínez, and I.G. Kevrekidis.
\newblock Identification of distributed parameter systems: A neural net based
  approach.
\newblock {\em Computers \& Chemical Engineering}, 22:S965--S968, 1998.
\newblock European Symposium on Computer Aided Process Engineering-8.

\bibitem{survey_kf_ann}
Ashesh Chattopadhyay, Pedram Hassanzadeh, Krishna~V. Palem, and Devika
  Subramanian.
\newblock Data-driven prediction of a multi-scale lorenz 96 chaotic system
  using a hierarchy of deep learning methods: Reservoir computing, ann, and
  {RNN-LSTM}.
\newblock {\em CoRR}, abs/1906.08829, 2019.

\bibitem{Sindy}
Steven~L. Brunton, Joshua~L. Proctor, and J.~Nathan Kutz.
\newblock Discovering governing equations from data by sparse identification of
  nonlinear dynamical systems.
\newblock {\em Proceedings of the National Academy of Sciences},
  113(15):3932--3937, 2016.

\bibitem{jaideep1}
Jaideep Pathak, Zhixin Lu, Brian~R. Hunt, Michelle Girvan, and Edward Ott.
\newblock Using machine learning to replicate chaotic attractors and calculate
  lyapunov exponents from data.
\newblock {\em Chaos: An Interdisciplinary Journal of Nonlinear Science},
  27(12):121102, 2017.

\bibitem{nielsen2019practical}
A.~Nielsen.
\newblock {\em Practical Time Series Analysis: Prediction with Statistics and
  Machine Learning}.
\newblock O'Reilly Media, 2019.

\bibitem{kaptanoglu2021physicsconstrained}
Alan~A. Kaptanoglu, Kyle~D. Morgan, Chris~J. Hansen, and Steven~L. Brunton.
\newblock Physics-constrained, low-dimensional models for magnetohydrodynamics:
  {{First-principles}} and data-driven approaches.
\newblock {\em Physical Review E}, 104(1):015206, July 2021.

\bibitem{kutz2022parsimony}
J.~Nathan Kutz and Steven~L. Brunton.
\newblock Parsimony as the ultimate regularizer for physics-informed machine
  learning.
\newblock {\em Nonlinear Dynamics}, January 2022.

\bibitem{bhcm1}
B.Haasdonk {,}B.Hamzi {,} G.Santin~{,} D.Wittwar.
\newblock Kernel methods for center manifold approximation and a weak
  data-based version of the center manifold theorems.
\newblock {\em Physica D}, 2021.

\bibitem{lyap_bh}
P.~Giesl, B.~Hamzi, M.~Rasmussen, and K.~Webster.
\newblock Approximation of {L}yapunov functions from noisy data.
\newblock {\em Journal of Computational Dynamics}, 2019.
\newblock \url{https://arxiv.org/abs/1601.01568}.

\bibitem{BHPhysicaD}
Boumediene Hamzi and Houman Owhadi.
\newblock Learning dynamical systems from data: A simple cross-validation
  perspective, part i: Parametric kernel flows.
\newblock {\em Physica D: Nonlinear Phenomena}, 421:132817, 2021.

\bibitem{hamzi2019kernel}
Boumediene Hamzi and Fritz Colonius.
\newblock Kernel methods for the approximation of discrete-time linear
  autonomous and control systems.
\newblock {\em SN Applied Sciences}, 1(7):674, July 2019.

\bibitem{bh2020b}
Stefan Klus, Feliks Nuske, and Boumediene Hamzi.
\newblock Kernel-based approximation of the koopman generator and
  schr{\"o}dinger operator.
\newblock {\em Entropy}, 22, 2020.
\newblock \url{https://www.mdpi.com/1099-4300/22/7/722}.

\bibitem{klus2020data}
Stefan Klus, Feliks N{\"u}ske, Sebastian Peitz, Jan-Hendrik Niemann, Cecilia
  Clementi, and Christof Sch{\"u}tte.
\newblock Data-driven approximation of the koopman generator: Model reduction,
  system identification, and control.
\newblock {\em Physica D: Nonlinear Phenomena}, 406:132416, 2020.

\bibitem{ALEXANDER2020132520}
Romeo Alexander and Dimitrios Giannakis.
\newblock Operator-theoretic framework for forecasting nonlinear time series
  with kernel analog techniques.
\newblock {\em Physica D: Nonlinear Phenomena}, 409:132520, 2020.

\bibitem{bhks}
Andreas Bittracher{,} Stefan Klus{,} Boumediene Hamzi{,}~Peter Koltai{,} and
  Christof Schutte.
\newblock Dimensionality reduction of complex metastable systems via kernel
  embeddings of transition manifold, 2019.
\newblock https://arxiv.org/abs/1904.08622.

\bibitem{bh12}
Jake Bouvrie and Boumediene Hamzi.
\newblock Empirical estimators for stochastically forced nonlinear systems:
  Observability, controllability and the invariant measure.
\newblock {\em Proc. of the 2012 American Control Conference}, pages 294--301,
  2012.
\newblock \url{https://arxiv.org/abs/1204.0563v1}.

\bibitem{bh17}
Jake Bouvrie and Boumediene Hamzi.
\newblock Kernel methods for the approximation of nonlinear systems.
\newblock {\em SIAM J. Control and Optimization}, 2017.
\newblock \url{https://arxiv.org/abs/1108.2903}.

\bibitem{hb17}
Jake Bouvrie and Boumediene Hamzi.
\newblock Kernel methods for the approximation of some key quantities of
  nonlinear systems.
\newblock {\em Journal of Computational Dynamics}, 1, 2017.
\newblock \url{http://arxiv.org/abs/1204.0563}.

\bibitem{mmd_kernels_bh}
Boumediene Hamzi, Christian Kuehn, and Sameh Mohamed.
\newblock A note on kernel methods for multiscale systems with critical
  transitions.
\newblock {\em Mathematical Methods in the Applied Sciences}, 42(3):907--917,
  2019.

\bibitem{MacKay2003}
David J.~C. MacKay.
\newblock {\em Information Theory, Inference, and Learning Algorithms}.
\newblock Copyright Cambridge University Press, 2003.

\bibitem{solomonoff1960preliminary}
R.~J. Solomonoff.
\newblock A preliminary report on a general theory of inductive inference
  (revision of report v-131).
\newblock {\em Contract AF}, 49(639):376, 1960.

\bibitem{kolmogorov1965three}
A.N. Kolmogorov.
\newblock Three approaches to the quantitative definition of information.
\newblock {\em Problems of information transmission}, 1(1):1--7, 1965.

\bibitem{chaitin1975theory}
Gregory~J Chaitin.
\newblock A theory of program size formally identical to information theory.
\newblock {\em Journal of the ACM (JACM)}, 22(3):329--340, 1975.

\bibitem{solomonoff1964formal}
Ray~J Solomonoff.
\newblock A formal theory of inductive inference. part i.
\newblock {\em Information and control}, 7(1):1--22, 1964.

\bibitem{levin1974laws}
L.A. Levin.
\newblock Laws of information conservation (nongrowth) and aspects of the
  foundation of probability theory.
\newblock {\em Problemy Peredachi Informatsii}, 10(3):30--35, 1974.

\bibitem{hutter2007algorithmic}
Marcus Hutter, Shane Legg, and Paul~MB Vitanyi.
\newblock Algorithmic probability.
\newblock {\em Scholarpedia}, 2(8):2572, 2007.

\bibitem{dingle2024exploring}
Kamaludin Dingle, Mohammad Alaskandarani, Boumediene Hamzi, and Ard~A Louis.
\newblock Exploring simplicity bias in 1d dynamical systems.
\newblock {\em arXiv preprint arXiv:2403.06989}, 2024.

\bibitem{arnold1995random}
Ludwig Arnold, Christopher~KRT Jones, Konstantin Mischaikow, Genevi{\`e}ve
  Raugel, and Ludwig Arnold.
\newblock {\em Random dynamical systems}.
\newblock Springer, 1995.

\bibitem{dingle2012knudsen}
Kamaludin Dingle, Jeroen~SW Lamb, and Joan-Andreu L{\'a}zaro-Cam{\'\i}.
\newblock Knudsen's law and random billiards in irrational triangles.
\newblock {\em Nonlinearity}, 26(2):369, 2012.

\bibitem{mayer1981influence}
G~Mayer-Kress and H~Haken.
\newblock The influence of noise on the logistic model.
\newblock {\em Journal of Statistical Physics}, 26:149--171, 1981.

\bibitem{dingle2018input}
Kamaludin Dingle, Chico~Q Camargo, and Ard~A Louis.
\newblock Input--output maps are strongly biased towards simple outputs.
\newblock {\em Nature communications}, 9(1):761, 2018.

\bibitem{dingle2020generic}
Kamaludin Dingle, Guillermo~Valle P{\'e}rez, and Ard~A Louis.
\newblock Generic predictions of output probability based on complexities of
  inputs and outputs.
\newblock {\em Scientific reports}, 10(1):1--9, 2020.

\bibitem{valle2018deep}
Guillermo Valle-Perez, Chico~Q Camargo, and Ard~A Louis.
\newblock Deep learning generalizes because the parameter-function map is
  biased towards simple functions.
\newblock {\em arXiv preprint arXiv:1805.08522}, 2018.

\bibitem{mingard2019neural}
Chris Mingard, Joar Skalse, Guillermo Valle-P{\'e}rez, David
  Mart{\'\i}nez-Rubio, Vladimir Mikulik, and Ard~A Louis.
\newblock Neural networks are a priori biased towards boolean functions with
  low entropy.
\newblock {\em arXiv preprint arXiv:1909.11522}, 2019.

\bibitem{bhattamishra2022simplicity}
Satwik Bhattamishra, Arkil Patel, Varun Kanade, and Phil Blunsom.
\newblock Simplicity bias in transformers and their ability to learn sparse
  boolean functions.
\newblock {\em arXiv preprint arXiv:2211.12316}, 2022.

\bibitem{yang2019fine}
Greg Yang and Hadi Salman.
\newblock A fine-grained spectral perspective on neural networks.
\newblock {\em arXiv preprint arXiv:1907.10599}, 2019.

\bibitem{hernandez2021algorithmic}
Santiago Hern{\'a}ndez-Orozco, Hector Zenil, J{\"u}rgen Riedel, Adam Uccello,
  Narsis~A Kiani, and Jesper Tegn{\'e}r.
\newblock Algorithmic probability-guided machine learning on non-differentiable
  spaces.
\newblock {\em Frontiers in artificial intelligence}, 3:567356, 2021.

\bibitem{dingle2023multiclass}
Kamaludin Dingle, Pau Batlle, and Houman Owhadi.
\newblock Multiclass classification utilising an estimated algorithmic
  probability prior.
\newblock {\em Physica D: Nonlinear Phenomena}, 448:133713, 2023.

\bibitem{mingard2023deep}
Chris Mingard, Henry Rees, Guillermo Valle-P{\'e}rez, and Ard~A Louis.
\newblock Do deep neural networks have an inbuilt occam's razor?
\newblock {\em arXiv preprint arXiv:2304.06670}, 2023.

\bibitem{li2008introduction}
M.~Li and P.M.B. Vitanyi.
\newblock {\em An introduction to Kolmogorov complexity and its applications}.
\newblock Springer-Verlag New York Inc, 2008.

\bibitem{doan2018hopf}
Thai~Son Doan, Maximilian Engel, Jeroen~SW Lamb, and Martin Rasmussen.
\newblock Hopf bifurcation with additive noise.
\newblock {\em Nonlinearity}, 31(10):4567, 2018.

\bibitem{zenil2011algorithmic}
Hector Zenil and Jean-Paul Delahaye.
\newblock On the algorithmic nature of the world.
\newblock In {\em Information and Computation: Essays on Scientific and
  Philosophical Understanding of Foundations of Information and Computation},
  pages 477--496. World Scientific, 2011.

\bibitem{dingle2023note}
Kamaludin Dingle, Rafiq Kamal, and Boumediene Hamzi.
\newblock A note on a priori forecasting and simplicity bias in time series.
\newblock {\em Physica A: Statistical Mechanics and Its Applications},
  609:128339, 2023.

\bibitem{Allen2014}
Kelty~Ann Allen.
\newblock {\em Martin-L\"of Randomness and Brownian Motion}.
\newblock PhD thesis, University of California, Berkeley, Berkeley, May 2014.
\newblock Available at \url{https://escholarship.org/uc/item/20072582}.

\bibitem{kanso2009logistic}
Ali Kanso and Nejib Smaoui.
\newblock Logistic chaotic maps for binary numbers generations.
\newblock {\em Chaos, Solitons \& Fractals}, 40(5):2557--2568, 2009.

\bibitem{calude2002information}
C.S. Calude.
\newblock {\em Information and randomness: An algorithmic perspective}.
\newblock Springer, 2002.

\bibitem{gacs1988lecture}
P.~G{\'a}cs.
\newblock {\em Lecture notes on descriptional complexity and randomness}.
\newblock Boston University, Graduate School of Arts and Sciences, Computer
  Science Department, 1988.

\bibitem{shen2022kolmogorov}
Alexander Shen, Vladimir~A Uspensky, and Nikolay Vereshchagin.
\newblock {\em Kolmogorov complexity and algorithmic randomness}, volume 220.
\newblock American Mathematical Society, 2022.

\bibitem{zenil2014correlation}
Hector Zenil, Fernando Soler-Toscano, Kamaludin Dingle, and Ard~A Louis.
\newblock Correlation of automorphism group size and topological properties
  with program-size complexity evaluations of graphs and complex networks.
\newblock {\em Physica A: Statistical Mechanics and its Applications},
  404:341--358, 2014.

\bibitem{dingle2022predicting}
Kamaludin Dingle, Javor~K Novev, Sebastian~E Ahnert, and Ard~A Louis.
\newblock Predicting phenotype transition probabilities via conditional
  algorithmic probability approximations.
\newblock {\em Journal of the Royal Society Interface}, 19(197):20220694, 2022.

\bibitem{alaskandarani2023low}
Mohammad Alaskandarani and Kamaludin Dingle.
\newblock Low complexity, low probability patterns and consequences for
  algorithmic probability applications.
\newblock {\em Complexity}, 2023, 2023.

\bibitem{johnston2022symmetry}
Iain~G Johnston, Kamaludin Dingle, Sam~F Greenbury, Chico~Q Camargo,
  Jonathan~PK Doye, Sebastian~E Ahnert, and Ard~A Louis.
\newblock Symmetry and simplicity spontaneously emerge from the algorithmic
  nature of evolution.
\newblock {\em Proceedings of the National Academy of Sciences},
  119(11):e2113883119, 2022.

\bibitem{sato2018dynamical}
Yuzuru Sato, Thai~Son Doan, Jeroen~SW Lamb, and Martin Rasmussen.
\newblock Dynamical characterization of stochastic bifurcations in a random
  logistic map.
\newblock {\em arXiv preprint arXiv:1811.03994}, 2018.

\bibitem{lind1995introduction}
Douglas Lind and Brian Marcus.
\newblock {\em An introduction to symbolic dynamics and coding}.
\newblock Cambridge university press, 1995.

\bibitem{kaspar1987easily}
F.~Kaspar and HG~Schuster.
\newblock Easily calculable measure for the complexity of spatiotemporal
  patterns.
\newblock {\em Physical Review A}, 36(2):842, 1987.

\bibitem{lempel1976complexity}
A.~Lempel and J.~Ziv.
\newblock On the complexity of finite sequences.
\newblock {\em Information Theory, IEEE Transactions on}, 22(1):75--81, 1976.

\bibitem{hasselblatt2003first}
Boris Hasselblatt and Anatole Katok.
\newblock {\em A first course in dynamics: with a panorama of recent
  developments}.
\newblock Cambridge University Press, 2003.

\bibitem{berger2001chaos}
Arno Berger.
\newblock {\em Chaos and Chance: An Introduction to Stochastic Apects of
  Dynamics}.
\newblock Walter de Gruyter, 2001.

\bibitem{crutchfield1982fluctuations}
James~Patrick Crutchfield, J~Doyne Farmer, and Bernardo~A Huberman.
\newblock Fluctuations and simple chaotic dynamics.
\newblock {\em Physics Reports}, 92(2):45--82, 1982.

\bibitem{hutter2004universal}
Marcus Hutter.
\newblock {\em Universal artificial intelligence: Sequential decisions based on
  algorithmic probability}.
\newblock Springer Science \& Business Media, 2004.

\bibitem{neth2023dilemma}
Sven Neth.
\newblock A dilemma for solomonoff prediction.
\newblock {\em Philosophy of Science}, 90(2):288--306, 2023.

\bibitem{hutter2007universal}
Marcus Hutter.
\newblock On universal prediction and bayesian confirmation.
\newblock {\em Theoretical Computer Science}, 384(1):33--48, 2007.

\bibitem{zenil2019algorithmic}
Hector Zenil, Narsis~A Kiani, Francesco Marabita, Yue Deng, Szabolcs Elias,
  Angelika Schmidt, Gordon Ball, and Jesper Tegner.
\newblock An algorithmic information calculus for causal discovery and
  reprogramming systems.
\newblock {\em Iscience}, 19:1160--1172, 2019.

\bibitem{solomonoff1978complexity}
Ray Solomonoff.
\newblock Complexity-based induction systems: comparisons and convergence
  theorems.
\newblock {\em IEEE transactions on Information Theory}, 24(4):422--432, 1978.

\bibitem{vitanyi2013similarity}
Paul~MB Vit{\'a}nyi.
\newblock Similarity and denoising.
\newblock {\em Philosophical Transactions of the Royal Society A: Mathematical,
  Physical and Engineering Sciences}, 371(1984):20120091, 2013.

\bibitem{zenil2019coding}
Hector Zenil, Liliana Badillo, Santiago Hern{\'a}ndez-Orozco, and Francisco
  Hern{\'a}ndez-Quiroz.
\newblock Coding-theorem like behaviour and emergence of the universal
  distribution from resource-bounded algorithmic probability.
\newblock {\em International Journal of Parallel, Emergent and Distributed
  Systems}, 34(2):161--180, 2019.

\bibitem{adriaans2007learning}
Pieter Adriaans.
\newblock Learning as data compression.
\newblock In {\em Computation and Logic in the Real World: Third Conference on
  Computability in Europe, CiE 2007, Siena, Italy, June 18-23, 2007.
  Proceedings 3}, pages 11--24. Springer, 2007.

\bibitem{deletang2023language}
Gr{\'e}goire Del{\'e}tang, Anian Ruoss, Paul-Ambroise Duquenne, Elliot Catt,
  Tim Genewein, Christopher Mattern, Jordi Grau-Moya, Li~Kevin Wenliang,
  Matthew Aitchison, Laurent Orseau, et~al.
\newblock Language modeling is compression.
\newblock {\em arXiv preprint arXiv:2309.10668}, 2023.

\end{thebibliography}



\appendix

\section{Estimating pattern complexity} 

To estimate complexity, we follow Ref.\ \cite{dingle2018input}  and use
\begin{equation}
C_{LZ}(x) =\begin{cases}
     \log_2(n), &  \hspace*{-0.3cm}  \text{$x=0^n$ or $1^n$}\\
    \log_2(n) [N_w(x_1...x_n) + N_{w}(x_n...x_1)]/2, & \hspace*{-0.2cm} \text{otherwise}
  \end{cases}\label{eq:CLZ}
\end{equation}
where $N_w(x)$ comes from the 1976 Lempel and Ziv complexity measure  \cite{lempel1976complexity}, and 
where the simplest strings $0^n$ and $1^n$ are separated out because  $ N_{w}(x)$ assigns complexity $K=1$ to the string 0 or 1, but complexity 2 to $0^n$ or $1^n$ for $n\geq2$, whereas the true Kolmogorov complexity of such a trivial string actually scales as $\log_2(n)$ for typical $n$ because one only needs to encode $n$. 
Having said that, the minimum possible value is $K(x)\approx0$ for a simple set, and so e.g. for binary strings of length $n$, we can expect $0 \leq K(x) \leq n$ bits. Because for a random string of length $n$ the value $C_{LZ}(x)$ is often much larger than $n$, especially for short strings, we scale the complexity so that $a$ in Eq.\ (\ref{eq:simplicity_bias}) is set to $a=1$ via
\begin{equation}
\tilde{K}(x) = \log_2(M) \cdot \frac{ C_{LZ}(x) - \min_x (C_{LZ})}{\max_x (C_{LZ}) - \min_x (C_{LZ}) } \label{eq:Kscaled}
\end{equation}
where $M$ is the maximum possible number of output patterns in the system, and the min and max complexities are over all strings $x$ which the map can generate.  $\tilde{K}(x)$ is the approximation to Kolmogorov complexity that we use throughout. 
This scaling results in $0\leq \tilde{K}(x) \leq n$ which is the desirable range of values.

\section{Comparison of Laplace and algorithmic probability for induction}

Here we give another example in which Laplace's induction method gives different conclusions as compared to induction based on algorithmic probability.

In the logistic map, assume that $\mu=2.5$ and $\epsilon=0.0$. As in the main text, if we assume that $x_0\approx 0$, then the digitised trajectory will begin as zeros, and then change to 1s as the trajectory approaches $1-1/\mu=$ 0.6$>$0.5. Define $x_0$ to be the value such that the series of 0s changes to 1s after $m^m$ zeros are observed, where $m$ is some positive integer. Because the logistic map can be described in a few bits, and $\mu=2.5$ and $\epsilon=0.0$ can also be described in a few bits, then the largest contribution to the complexity of the series comes from the value of $m$. We have
\begin{equation}
  K(m^m) = K(m) +O(1) \approx \log_2(m)
\end{equation}
for typical and large $m$. Therefore the algorithmic probability prediction that the next bit after the zeroes will be 1 is roughly $2^{-\log_2(m)}\sim 1/m$. In contrast, after observing $m^m$ zeros, the Laplacian prediction for the probability that the next bit is 1 is $1/m^m$ which is substantially smaller. Therefore the observation of a 1 is more `surprising' from the perspective of the Laplacian method.

\end{document}